\documentclass[12pt]{article}
\usepackage{sc3conf} 
\usepackage{amsfonts} 
\usepackage{amsmath} 

\def\BRST{(\ref{BRST})} 
\def\BRSTA{(\ref{BRSTA})} 
\def\BRSTB{(\ref{BRSTB})}

\def\GSC{(\ref{GSC})}

\def\YMa{(\ref{YMa})}

\def\YMc{(\ref{YMc})}

\def\nilB{(\ref{nilB})}

\def\nilE{(\ref{nilE})}

\def\puuI{(\ref{puuI})} 
 
\def\susyA{(\ref{susyA})} 
\def\susyB{(\ref{susyB})} 
\def\susyC{(\ref{susyC})} 
 
\newcommand{\ahin}[2]{\begin{eqnarray}\label{#1}{#2}\end{eqnarray}} 

\newcommand{\foot}[1]{\footnote{#1}}

\def\semi{;\\ }

\def\GreenSchwarz{\cite{GreenSchwarz}} 
 
\def\superstring{\cite{superstring}}

\def\GrassiUG{\cite{GrassiUG}} 
 
\def\GrassiSP{\cite{GrassiSP}} 
 
\def\GrassiXV{\cite{GrassiXV}}

\def\GrassiSR{\cite{GrassiSR}} 

\def\Grassi{\cite{GrassiUG,GrassiSP,GrassiXV,GrassiXF,GrassiSR}}

\def\tonin{\cite{tonin}}

\def\GrassiXF{\cite{GrassiXF}} 
 
\def\berko{\cite{berko}}

\def\homo{\cite{homo}}

\def\p{\partial}

\def\half {{\frac 1 2 }}

\def\a{\alpha} 
\def\b{\beta} 
\def\g{\gamma}  \def\G{\Gamma} 
\def\d{\delta}   
\def\m{\mu} 
\def\n{\nu} 
 
\def\l{\lambda}  
\def\k{\kappa} 
\def\e{\epsilon}

\def\a{\alpha} 
\def\b{\beta} 
\def\d{\delta} 
 
\def\m{\mu} 
\def\n{\nu} 
 
\def\s{\sigma} 
\def\l{\lambda}

\def\k{\kappa}

\def\t{\theta}

\def\|{\Big|} 
\def\({\Big(}   \def\){\Big)} 
\def\[{\Big[}   \def\]{\Big]} 
 
\def\a{\alpha} 
\def\b{\beta} 
\def\g{\gamma} 
\def\l{\lambda} 
\def\d{\delta} 
\def\e{\epsilon} 
\def\t{\theta} 
 
\def\s{\sigma}

\def\G{\Gamma}

\def\p{\partial} 
\def\half{{\frac 1 2 }}

\def\pg{{\rm pg}} 
\def\af{{\rm af}} 
 
\font\frtnfr=eufm10   scaled\magstep1 \font\twlfr=eufm10 
\font\tenfr=eufm10   
\newfam\frfam 
\textfont\frfam=\frtnfr \scriptfont\frfam=\twlfr 
\scriptscriptfont\frfam=\tenfr

\font\frtnopen=msbm10  scaled\magstep2 \font\twlopen=msbm10 
\font\tenopen=msbm10   
\newfam\openfam 
\textfont\openfam=\frtnopen \scriptfont\openfam=\twlopen 
\scriptscriptfont\openfam=\tenopen

\font\frtnsf = cmss12 scaled\magstep1 \font\twlsf = cmss10 
\font\tensf = cmss9   
\newfam\Scfam 
\textfont\Scfam = \frtnsf \scriptfont\Scfam = \twlsf 
\scriptscriptfont\Scfam = \tensf 
 
\renewcommand{\thefootnote}{\fnsymbol{footnote}} 
 
\begin{document} 
\raggedbottom 
 
\title{An Introduction to the \\ Covariant Quantization of Superstrings} 
 
\authors{P.A.~Grassi,\footnotemark{}\adref{1} 
  G.~Policastro,\footnotemark{}\adref{2} 
  and P.~van~Nieuwenhuizen\footnotemark{}\adref{1}} 
 
\addresses{\1ad  C.N. Yang Institute for Theoretical Physics, 
                  State University of New York\\ at Stony Brook, 
                  NY 11794-3840, USA 
  \nextaddress 
  \2ad  DAMTP, Centre for Mathematical Sciences 
               Wilberforce Road,\\ Cambridge CB3 0WA, UK} 
 
\maketitle 
 
\makeatletter 
\renewcommand{\@makefnmark}
{\mbox{$^{\fnsymbol{footnote}}$}} 
\makeatother 
\setcounter{page}{1}

\setcounter{footnote}{1} 
\footnotetext{pgrassi@insti.physics.sunysb.edu} 
\addtocounter{footnote}{1} 
\footnotetext{policast@cibslogin.sns.it} 
\addtocounter{footnote}{1} 
\footnotetext{vannieu@insti.physics.sunysb.edu}

\begin{abstract} 

  We give an introduction to a 
  new approach to the covariant quantization of superstrings. 
  After a brief review of the classical Green--Schwarz 
  superstring and Berkovits' approach to its quantization based on 
  pure spinors, we discuss our covariant formulation without pure spinor 
  constraints. We discuss the relation between the concept of grading, 
  which we introduced to define vertex operators, and 
  homological perturbation theory, and we compare our work with recent 
  work by others. In the appendices, 
  we include some background material for the Green-Schwarz and 
  Berkovits formulations, in order that this presentation be self 
  contained. 
\end{abstract} 
 
 
\renewcommand{\thefootnote}{\arabic{footnote}} 
\setcounter{footnote}{0} 
 
\makeatletter 
\renewcommand{\@makefnmark}{\mbox{$^{\@thefnmark}$}} 
\makeatother

\section{Introduction} 
String theory is mostly based on the Ramond--Neveu--Schwarz (RNS) 
formulation, with worldsheet fermions $\psi^m$ in the vector 
representation of the spacetime Lorentz group $SO(9,1)$. This 
formulation exhibits classically a $N=1$ local supersymmetry of the 
worldsheet.  The BRST symmetry of the RNS formulation is based on the 
super-reparametrization invariance of the worldsheet. The fundamental 
fields are the bosons $x^m$, the fermions $\psi^m$, the 
reparametrization ghosts $b_{zz}, c^z$ and the superghosts $\beta_{\small {z \sqrt{z}}}, 
\g^{\small \sqrt{z}}$. Physical states correspond to vertex operators which {i)} belong 
to the BRST cohomology and {ii)} are annihilated by the zero mode $b_0$ of the 
antighost for the open string, or by $b_0$ and $\tilde b_0$ for the closed string.  To obtain 
a set of physical states which form a representation of spacetime 
supersymmetry, the GSO projection is applied to remove half of the 
physical states. Spacetime supersymmetry is thus not manifest, and the 
study of Ramond-Ramond backgrounds is not feasible. Therefore, one 
would prefer a formulation with spacetime fermions $\t^\a$ belonging 
to a representation of $Spin(9,1)$ because it would keep spacetime 
supersymmetry (susy) manifest. 

At the classical level, such a 
formulation was constructed by Green and Schwarz in 1984 
\GreenSchwarz. Their classical action contains two fermions 
$\t^{i\a}$ ($i=1,2$) and the bosonic coordinates $x^m$. Each of the 
$\t$'s is real and can be chiral or anti-chiral (type IIA/B 
superstrings): they are $16$-component Majorana-Weyl spinors which are 
spacetime spinors and worldsheet scalars. We denote chiral 
spinors by contravariant indices $\t^\a$ with $\a=1,\dots,16$; 
antichiral spinors are denoted by $\t_\a$, also with $\a=1,\dots,16$. 
We shall only consider chiral $\t$'s below.  

The rigid spacetime supersymmetry is given by the usual non-linear 
coordinate representation 
\ahin{susyA}{\d_\e \t^{i\a} = \e^{i\a}\,, ~~~~~ 
\d_\e x^m = \bar\e^i \G^m \t^i = i\, \e^{i \a} \g^m_{\a\b} \t^{i \b}\,,} 
where $\g^m_{\a\b}$ are  ten real symmetric $16\times16$ matrices and the 
flavor indices $i=1,2$ are summed over. (In appendix A, Dirac matrices and 
Majorana-Weyl spinors are reviewed). 
Susy-invariant building blocks are 
\ahin{susyB}{\Pi^m_\mu \equiv \p_\mu x^m - 
i \t^i \g^m \p_\mu \t^i\,, ~~~~~ \p_\mu \t^{i\a}\,} where $\mu=0,1$ 
and $\p_0 = \p_t$ and $\p_1 = \p_\s$. A natural choice for the action 
on a flat background spacetime and curved worldsheet would seem to be 
\ahin{susyC}{{\cal L}_1 = -\frac {1}{ 2  \pi} \, \sqrt{-h} \, 
h^{\mu\nu} \eta_{mn} \, \Pi^m_\mu \Pi^n_\nu\,, } 
with $h^{\m\n}$ the worldsheet metric, 
because it is the susy-invariant line element (a natural generalization of the 
action for the bosonic string). However, it yields no kinetic term for the fermions. Even if 
one could produce a kinetic term, there would still be the problem that one would have 
${\frac 1 2 }(16+16) = 16$ fermionic propagating modes and $8$ bosonic 
propagating modes. Such a theory could not yield a linear representation of supersymmetry. 
 
A resolution of this problem became possible when Siegel found a new 
local fermionic symmetry ($\k$-symmetry) for the point particle 
\cite{SiegelHH}. 
Green and Schwarz tried to find this symmetry in their string, and they discovered 
that it is present, but only after adding a Wess-Zumino-Novikov-Witten 
term to the action. Using this symmetry one could impose the gauge 
$\G^+ \t^1 = \G^+ \t^2 =0$ (where $\G^\pm = \G^0 \pm \G^9$), and if 
one then also fixed the local scale and general coordinate symmetry by 
$h^{\mu \nu } = \eta^{\mu\nu}$, and the remaining conformal symmetry by 
$x^+(\s,t) = x^+_0 + p^+ t$, the action became a free string theory with $8$ 
fermionic degrees of freedom and $8$ bosonic degrees of freedom. Susy 
was linearly realized and quantization posed no problem. 
 
However, in this combined $\k$-light cone gauge, manifest $SO(9,1)$ 
Lorentz invariance is lost, and with it all the reasons for studying the 
superstring in the first place. (We shall call the string of 
Green and Schwarz the superstring, to distinguish it from the RNS 
string which we call the spinning string.) 
 
Going back to the original classical action, it was soon 
realized that second class constraints were present, due to the 
definition of the conjugate momenta of the $\t$'s. These second class 
constraints could be handled by decomposing them w.r.t. a non-compact 
$SU(5)$ subgroup of $SO(9,1)$ (see appendix D) , but then again 
manifest Lorentz invariance was lost. An approach to quantization 
which could deal with second class constraints and keep covariance 
was needed. By using a 
proposal of Faddeev and Fradkin, one could 
turn second class constraints into first class constraints by adding further fields, 
but upon quantization one now obtained an infinite set of ghosts-for-ghosts, and 
problems with the calculation of anomalies were encountered. At the end 
of the 80's, several authors tried different approaches, but they always 
encountered infinite sets of ghosts-for-ghosts, and 15 years of pain 
followed \superstring. 
 
A few years ago Berkovits developed a new line of thought \berko. 
Taking a flat background and a flat worldsheet metric, 
the central charge $c$ in one sector\footnote{One $\t$ is always left-moving and the other $\t$ is 
always right-moving, whether or not they have the same chirality.} 
of 10 free bosons $x^m$ and one $\t$ is 
$c =10 - 2 \times 16 =-22$ 
(there is a conjugate momentum $p_{z\a}$ for $\t^\a$). He noted 
that if one decomposes a chiral spinor $\l^\a$ under the 
non-compact $SU(5)$ subgroup of $SO(9,1)$, it decomposes as 
$\underline{16} \rightarrow 
\underline{10}+\underline{5}^*+\underline{1}$ (see Appendix~D). 
Imposing the constraint 
\ahin{pure}{\l^T \g^m \l = 0\,,} 
also known as {\it pure spinor constraint}, 
one can express the $\underline{5}^*$ in terms of the $\underline{10}$ 
and $\underline{1}$, and hence it seemed that by adding a commuting 
pure spinor (with conjugate momenta for the $\underline{10}$ 
and $\underline{1}$), one could obtain vanishing central charge: 
$c = 10_x - 2 \times 16_{\t,p_\t} + 2 \times (10 +1)_{\l, p_\l} = 0$. In 
the past few years, he has developed this approach further. 
 
Having a constraint such as (\ref{pure})~in a theory 
leads to problems at the quantum level in the computation of loop 
corrections and in the definition of the path integral. A similar situation 
occurred in superspace formulations of supergravity, where one must impose constraints 
on the supertorsions; in that case the constraints were solved and the 
covariance was sacrificed. One could work only with $\underline{10}$ and 
$\underline{1}$, but then one would again violate manifest Lorentz 
invariance. 
 
We have developed an approach 
\Grassi~which starts with the same 
$\t^\a, p_{z\a}$ 
and $\l^\a$ as used by Berkovits, but we relax the constraint (\ref{pure})~by 
adding new ghosts. In Berkovits' and our approach one has the BRST 
law $s\, \t^\a = i \l^\a$, with real $\t^\a$, but in Berkovits' approach $\l^\a$ 
must be complex in order that (\ref{pure})~have a solution at all, 
{\it whereas in our approach $\l^\a$ is real}. 
The law $s \t^\a = i\, \l^\a$ is an enormous simplification over the law one would 
obtain from the $\kappa$-symmetry law $\delta_\kappa \t^\a = \Pi^m_\mu (\g_m \k^\mu)^\a$ 
with selfdual $\k^{\mu}_\beta$. It is this simpler starting point that avoids the 
infinite set of ghosts-for-ghosts. Perhaps our finite set of fields corresponds to 
resummations of the infinite set of fields 
encountered in previous approaches. 
First, we give a brief review of the classical superstring action from which 
we shall only extract a set of first class constraints $d_{z\a}$. These first 
class constraints are removed from the action and used to construct a BRST charge. 
 
We deduce the full theory 
by requiring nilpotency of the BRST charge: each time nilpotency on a given field 
does not hold we add a new field (ghost) and define its BRST transformation 
rule such that nilpotency holds.  A priori, one might expect that one would end up again with an 
infinite set of ghosts-for-ghosts, but to our happy surprise the iteration procedure 
stops after a finite number of steps. 
 
In some modern approaches the difference between the action and the BRST charge becomes 
less clear (in the BV formalism the action is even equal to the BRST charge, and 
in string field theory the action is expressed in terms of the BRST charge). So the 
transplantation of the first class constraints from the action to the BRST charge may not be as 
drastic as it may sound at first. We may in this way create a different off-shell formulation of the 
same physical theory. The great advantage of this procedure is that one is left with a free 
action, so that propagators become very easy to write down, and OPE's among vertex operators 
become as easy as in the RNS approach. 
 
We shall now present our approach. We have a new definition of physical states, 
and we obtain the correct spectrum for the open string as well as for the closed 
superstring, both at the massless level and at the massive levels. Since these 
notes are intended as introduction to our work, we give much background material 
in the appendices. Such material is not present in our papers, but 
may help to understand the reasons and the technical aspects of our approach. 
Our approach differs from the conventional approaches to the BRST quantization of strings. 
One would therefore like to see it work in a simpler example. For that reason we have applied 
our ideas to pure Yang-Mills field theory \GrassiSR. One gets also in that case the 
correct cohomology. 

We have found since the conference some deep geometrical meanings of the new ghosts, but we have 
not yet found the underlying classical action to which our quantum theory corresponds. Sorokin, Tonin and 
collaborators have recently shown \tonin~how one can obtain Berkovits' theory 
from a $N=(2,0)$ worldsheet action with superdiffeomorphism embeddings, and it is possible 
that a similar approach yields our theory. 
 
\section{The classical Green-Schwarz action} 
 
As we already mentioned, a natural generalization of the bosonic string with ${\cal L} \sim (\p_\a x^m)^2$ with 
spacetime supersymmetry is the supersymmetric line element given in \susyB~and \susyC. 
If one considers the interaction term $\p_\mu x^m (\t \g_m \p^\m \t)$ and if 
one chooses the light cone gauge $x^+ = x^+_0 + p^+ t$ one obtains a term 
$p^+ \t \g_+ \p_t \t= (\sqrt{p^+} \t) \g_+ \p_t (\sqrt{p^+} \t)$. This is not a satisfactory 
kinetic term because we also would need a term with $ p^+ \t \g_+ \p_\s \t$. Such a term 
would be obtained if the action contains a term of the form $ (\p_t x^+) \t \g_+ \p_\s \t$, or in covariant 
notation $\e^{\m\n} (\p_\m x^m) \t \g_m \p_\n \t$. 
The extra kinetic term  $\e^{\m\n} \p_\m x^m \t \g_m \p_\n \t$ is part of a Wess-Zumino term 
(see appendix B). 
 
Rigid susy \susyA~and $\d_\e (\p_\s x^+) = 0$ 
would lead to $\e \g^+ \p_\s \t =0$. This suggests that the light-cone gauge for $\t$ 
should read $\g^+ \t =0$, or, in terms of $32 \times 32$ matrices, $\Gamma^+ \t =0$. 
If $\Gamma^+ \t =0$, also $\t^T C \Gamma^+=0$, and inserting 
$\{\Gamma^+, \Gamma^-\} = 1$, one would also 
find that $\t^T C \Gamma^I \p_\s \t =0$ for $I=1, \dots, 8$. So, then we would find in the light cone gauge 
that the action for $\t$ becomes a free action, a good starting point for string theory at the 
quantum level. 
 
In order that these steps are correct, we would need a local fermionic symmetry which would justify 
the gauge $\Gamma^+ \t =0$.  Pursuing this line of thought, one arrives then at the 
crucial question: does the sum of the supersymmetric line element and the WZNW term contain a new fermionic 
symmetry with half as many parameters as there are $\t$ components ? The answer is affirmative, and the 
$\k$-symmetry is briefly discussed at the end of appendix B, but since we shall not need the explicit form of 
the $\kappa$ symmetry transformation laws, we do not give them. 
 
The superstring action is very complicated already in a flat background. We extract from it a set of 
first class constraints $d_{z\a}= 0$, from which we build the BRST charge, and at all stages we work with a 
free action. 
The precise way to obtain $d_{z\a}$ from the classical superstring action is discussed in appendix~C. 
 
 
\section{Determining the theory from  the nilpotency of the BRST charge} 
 
We now start our program of determining the theory 
(the BRST charge and the ghost content) by requiring 
nilpotency of the BRST transformations. We consider one $\t$ for 
simplicity (we have also extended 
our work to two $\t$'s \GrassiXF). 
We shall be careful (for once) with aspects such as reality and normalizations. 
The BRST transformations preserve reality and are generated by $\Lambda Q$ where $\Lambda$ 
is imaginary and anti-commuting. It then follows that $Q$ should also be antihermitian in order that 
$\Lambda\, Q$ be antihermitian. For any field, we define 
the $s$ transformations as BRST transformations without $\Lambda$, 
so $\d_B \Phi = [\Lambda Q, \Phi]$ and $s \Phi = [Q, \Phi]_{\pm}$. 
The $s$-transformations have reality properties 
which follow from the BRST transformations (which preserve reality). 
 
We begin with 
\ahin{BRSTA}{Q = \int i \l^\a d_{z\a}\,,} 
where $d_{z\a}$ is given in Appendix C and 
$\int = {\frac {1}{ 2 i \pi}} \oint dz$. This Q is indeed antihermitian because $d_{z \a}$ is 
antihermitian. (We have performed a Wick rotation in appendix C, in order to be able to use the conventional tools 
of conformal field theory, but the reality properties hold in Minkowski space). 
The BRST operator depends on Heisenberg fields which satisfy the field 
equations, and since we work with a free action, $\bar\p \l^\a =0$ and $\bar\p d_{z\a} =0$ so that 
in flat space $\l^\a d_{z \a}$ is a holomorphic current, namely 
$\bar\p (\l^\a d_{z \a}) = 0$. 
 
The field $d_{z\a}$ contains a term $p_{z\a}$, where $p_{z\a}$ is the 
momentum conjugate to $\t^\a$ and it is antihermitian since $p_{z\a}$ 
is antihermitian as can be seen from the action $\int d^2z p_{z\a} \bar\p \t$. 
The factor ${\frac 1 2}$ in  $d_{z\a}$ in eq. \GSC\ 
can be checked by noting that the OPE\foot{The OPE of $d_\a$ with $d_\b$ is evaluated 
using $\p x^m(z) \p x^n(w) \sim - \eta^{mn} (z-w)^{-2}$ and $p_{z\a}(z) \t^\b (w) \sim \d_\a^{~~\b} (z-w)^{-1}.$} 
of $d_\a$ with $d_\b$ be proportional to $\Pi^m_w$. The expression for $\Pi^m_z$ is real and fixed by spacetime 
susy. 
 
The operators $d_{z\a}$ generate a closed algebra of currents with a central charge 
\ahin{BRSTB}{ 
d_\a(z) d_\b(w) \sim  2 i {\frac{\g^m_{\a\b}\Pi_m(w)}{z-w}}\,, 
~~~~~~~ 
d_\a(z) \Pi^m(w) \sim - 2 i {\frac{\g^m_{\a\b}\p\t^\b(w)} {z-w}}\,, 
} 
$$ 
\Pi^m(z) \Pi^n(w) \sim - \frac{ 1}{(z-w)^2} \eta^{mn}\,,~~~~~~~ 
    d_\a(z) \t^\b(w) \sim {\frac 1 {z-w}} \delta^{~\b}_{\a}\,. 
$$ 
 
Acting with \BRSTA~on $\t^\a$, one obtains 
$s \t^\a = i \l^\a$, and acting on $\l^\a$  yields $s \l^\a =0$. Nilpotency 
on $\t^\a$ and $\l^\a$ is achieved. 
Repeating this procedure on $x^m$ 
gives $s x^m = \l \g^m \t$, but since $s^2 x^m = i \l \g^m \l$ 
does not vanish, we introduce a new real anticommuting ghost $\xi^m$ by setting 
$s\,  x^m = \l \g^m \t + \xi^m$ and choosing the BRST transformation law of 
$\xi^m$ such that the nilpotency on $x^m$ is obtained. This leads to 
$s \xi^m = - i \l \g^m\l$. Nilpotency on $x^m$ is now achieved, but $s$ 
has acquired an extra 
term\foot{Spacetime susy requires that $Q'$ depends on $\Pi^m_z$ 
instead of, for example $\p_z x^m$.} 
$Q' = - \int \xi^m \Pi_{z m}$ where we recall $\Pi^m_z  = 
\p_z x^m - i \t \g^m \p_z \t$. Nilpotency on $p_{z\a}$, or 
equivalently on $d_{z\a}$, is obtained by further 
modifying the sum of 
$Q d_{z\a} = 
- 2 \Pi^m_z (\g_m \l)_\a$ and 
$ Q' d_{z \a} = - 2 i \xi^m (\g_m \p_z \t)_\a$ 
by adding 
$Q''  d_{z \a} = 
\p_z\chi_\a$ and fixing the BRST law of $\chi_\a$ such that nilpotency on 
$d_{z\a}$ is achieved.\foot{Since $(Q+Q')^2 d_{z\a} = \p_z( -2 \xi^m \g_m \l) _\a$, 
we add a term $\p_z \chi_\a$ instead of a field $\chi_{z\a}$.} 
This yields $Q \chi_\a = 2\xi^m (\g_m \l)_\a$ 
and $Q^2 \chi_\a =0$ due to a Fierz rearrangement involving three 
chiral spinors (see eq.~(\ref{ccccc})). At this point we have achieved nilpotency on $\t^\a, 
x^m, d_{z\a}$ and $\l^\a, \xi^m, \chi_\a$. We introduce the antighosts 
$w_{z\a}, \b_{z m}, \k^\a_z$ for the ghosts $\l^\a, \xi^m, \chi_\a$ and 
find that $s \Phi = [Q, \Phi\}$ with 
\ahin{BRST}{ 
Q = \oint \Big(i\l^\a d_{z\a} -\xi^m \Pi_{z m} - \chi_\a \p_z \t^\a 
- 2 \xi^m (\k_z \g_m \l) -i \b_{z m} \l\g^m \l\Big) 
} 
reproduces all BRST laws for all fields introduced so far except for 
the three antighosts.
 
Unfortunately, the BRST charge \BRST~fails to be nilpotent and 
therefore the concept of BRST cohomology is at this point 
meaningless. In order to repair this problem, we could proceed in two different 
ways: {\it i)} either continuing with our program of requiring nilpotency on each field 
separately (continuing with the antighosts $\beta_{zm}$, $\kappa_z^\a$ and $w_{z\a}$); or {\it ii)} 
terminate this process by hand in one stroke by adding a ghost 
pair $(b,c_z)$ as we now explain. 
We begin with 
\ahin{nilA}{ 
\{Q,Q\} = \int A_z \,,~~~~~~~ A_z = \xi_m \partial_z \xi^m + i  \lambda^\a \partial_z \chi_\a 
- i \chi_\a \partial_z \lambda^\a \,.} 
The non-closure term $A_z$ is due to the double poles 
in \BRSTB. By direct computation we establish that the anomaly $\int A_z$ is 
BRST invariant, as it should be according to consistency, 
$[Q,A_z] = \partial_z Y$ where $Y = {i} \xi_m \lambda \g^m \lambda$. 
If we define 
\ahin{nilB}{ 
Q' = Q + \int \left( c_z - b B_z\right)\,, 
} 
with an hermitian $c_z$ and an antihermitian $b$, 
we find that 
\ahin{nilC}{ \{Q',Q'\} = \int \Big( (A_z - B_z) +  b [Q,B_z] \Big)\,,} 
and, requiring that $Q'$ be nilpotent, a solution for $B_z$ is obtained by 
imposing\footnote{The relation $[Q,X] = - Y$ follows from acting with $Q$ on $B_z = A_z + \partial_z X$.}  
\ahin{nilD}{ 
[Q,B_z]=0 \,, ~~~~~ B_z = A_z + \partial_z X \,,~~~~~ [Q,X] = - Y 
} 
which is satisfied by $X=- {\frac i 2 } \chi_\a \lambda^\a$. Then one 
gets\footnote{One can even obtain a nilpotent current: $j' = j + c - b B + \partial( b X)$. 
Use that Q, but not $j$, commutes with $ -  \int b B$.}  
\ahin{nilE}{ 
B_z = \xi_m \partial_z \xi^m + {\frac i 2 } \lambda^\a \partial_z \chi_\a - {\frac{3i}{ 2}} 
\chi_\a \partial_z \lambda^\a\,. 
} 
 
However, any $Q'$ of the form $\int c_z +$``more'' can be always  brought in the form 
$\int c_z$ by a similarity transformation, 
namely as follows 
\ahin{newup}{ 
Q' = \Big[ e^{\int (- R_z - b\, S_z - b \p_z b\, T)} \int c_z 
e^{\int (R_z + b\, S_z + b \p_z b\, T)} 
\Big] 
} 
$$ 
=\int \left(c_z + S_z - b \p_z T \right) + 
\frac{1}{2}\Big[\int \left( S_z - b \p_z T \right), {\cal U} \Big] + 
{\frac {1}{6} }  \Big[ \Big[ \int \left(  S_z - b \p_z T \right), {\cal U} \Big],  {\cal U} \Big] + \dots 
$$ 
where $ {\cal U} = \int (R_z + b\, S_z + b \p_z b\, T)$. The $R_z, S_z$ and 
$T$ are hermitian polynomials in all fields except $c_z,b$ with ghost numbers 
$0,1,2$, respectively.  The solution in \nilB~and \nilE~corresponds to a particular 
choice of $R_z, S_z$ and $T$\footnote{Namely, $T = {1\over 2} X, \int S_z = Q, R_z =0$. All the 
terms displayed in (\ref{newup}) contribute.}, but any other choice also yields a nilpotent BRST 
charge. 

There is now a problem: 
the operator $Q' = e^{-{\cal U}} \int c_z  e^{{\cal U}}$ has trivial 
cohomology in the space of local vertex operators, 
because any ${\cal O}(w)$ satisfying ${\int c_z} {\cal O}(w) =0$ can always be written 
as ${\cal O}(w) = \int c_z {\cal G}(w)$ where ${\cal G}(w) =  b_0{\cal O}(w)$. 
(Note that ${\cal O}(w)$ cannot depend on $b_0$ because $\int c_z {\cal O}(w) =0$, and $c_0 = \int c_z$). 
 
We shall restrict 
the space of vertex operators in which $Q$ acts, in order to obtain non-trivial 
cohomology. We achieve this by introducing a new quantum number, called grading, 
and requiring that vertex operators have non-negative grading. In the 
smaller space of non-negative grading (see next section) the similarity transformation 
cannot transform each $Q$ into the form $\int c_z $, and we shall indeed 
obtain non-trivial cohomology, namely the correct cohomology. 

We have at this point obtained a new nilpotent BRST charge, and a set of ghost (and 
antighost) fields (whose geometrical meaning at this point is becoming clear). It is time 
to revert to the issue of the central charge. Since all fields are free 
fields, one simply needs to add the central charge of each canonical pair: this yields  
$c = 20$. So the central charge does not vanish, and to remedy this obstruction, we 
add by hand a real anticommuting vector pair $(\omega^m,\eta^m_z)$ which contributes 
$- 2 \times 10 $ to $c$. The BRST charge does not contain $\omega^m$ and $\eta^m_z$, 
hence  $\omega^m$ and $\eta^m_z$ are BRST inert. 
 
The reader (and the authors) may feel uncomfortable with these 
rescue missions by hand; a good theory should produce all fields automatically 
without outside help. Fortunately, we can announce that a more 
fundamental way of proceeding, by  continuing to require 
nilpotency on the antighosts and then on the new fields which are introduced 
 in this process, produces the pair $(\omega^m,\eta^m_z)$! We are in the process of writing 
these consideration up, and hopefully also the pair $(b,c_z)$ will be automatically 
produced in this way. 
 
Our results obtained by elementary methods and ad hoc additions, display nevertheless 
a few striking regularities, which confirm us in our belief that we are on the right track. 
For example, the grading which we discuss in the next section is generated by a current whose 
anomaly vanishes. This need not have happened, and provides welcome support 
for the various steps we have taken, but it hints of course at something more fundamental.
 
\section{The notion of the grading} 
 
In our work we define physical states by means of vertex operators which 
satisfy two conditions 
 
{\it i)} They are in the BRST cohomology 
 
{\it ii)} They should have non-negative grading \GrassiSP. 
 
The grading is a quantum number which was initially obtained from the algebra of the abstract currents 
$d_{z\a}, \Pi^m_z$ and $\p_z \t^\a$. Assigning grading $-1$ to $d_{z\a}$, we assign grading $+1$ 
to the corresponding ghost $\l^\a$. We then require that the grading be preserved in the 
operator product expansion. From $d d \sim \Pi$ we deduce that $\Pi^m_z$ has grading $-2$, 
so $\xi^m$ has grading $2$. Then $d \Pi \sim \p\t$ assigns grading $-3$ to $\p\t$, and thus grading $+3$ 
to $\chi$. (To avoid confusion note that in some of our pubblished work we use half these gradings).
The grading of the ghosts $b$ and $c$ is more subtle, but it can be obtained in the same 
spirit. From $d \p\t \sim (z-w)^{-2}$ and $\Pi \Pi \sim (z-w)^{-2}$ we introduce 
a central charge generator $I$ which has grading $-4$. The corresponding 
ghost $c_z$ has grading $4$. All antighosts have opposite grading from the ghosts. The trivial 
ghost pair $\omega^m, \eta^m_z$ has grading $(4,-4)$ because it is part of a quartet of which the grading 
of the other members is already known \GrassiSP. 
With these grading assignments to the ghost fields, the BRST charge can be 
decomposed into terms with definite but different gradings. It turns out that 
all the terms have non-negative grading: $Q = \sum_{n=0}^{4} Q_{n}$. 
This $Q$ maps the subspace of the Hilbert space with non-negative grading into itself. In 
\GrassiXV, the equivalence with Berkovits' pure spinor formulation has been 
proven. 
 
According to the grading condition {\it ii)}, the most general expression for the massless 
vertex in the case of open superstring is given by 
\ahin{graddA}{ 
{\cal O} = \l^\a A_\a + \xi^m A_m + \chi_\a W^\a + 
\omega_m B^m + b{\rm -terms}\, 
} 
where $A_\a, A_m, W^\a$ and $B^m$ are arbitrary superfields, so $A_\a = A_\a(x,\t)$, etc.. 
Requiring non-negative grading, the following combinations 
\ahin{gradB}{ 
b \l^\a \l^\b\,, ~~~~~~~ b \l^\a \xi^m\,, 
} 
are not allowed. Note that the vertex operator does not have a specific grading 
but contains terms with several (nonnegative) gradings. 

Finally, requiring BRST invariance of ${\cal O}$, one easily derives 
the equations of motion for $N=1$ SYM in $D=(9,1)$. From the $b$-terms in ${\cal O}$ 
one only finds that the superfields in these terms are expressed in terms of $A_\a, A_m, W^\a$ and $B^m$. 
However, in the sectors with $\l^\a \l^\b$ and $\l^\a \xi^m$ one learns that all remaining superfields 
appearing in this vertex operator can be expressed in terms of $A_\alpha(x,\t)$, for example
\ahin{bbbb}{
A_m = \frac{1}{8} \gamma^m_{\alpha \beta} D_\alpha A_\beta\,, 
~~~~~~~
W^\alpha = \frac{1}{10}  \gamma^m_{\alpha \beta} \left( D_\beta A_m - \partial_m A_\beta 
\right)\,.}
The superfield $A_\alpha$ itself satisfies 
\ahin{bbbc}{
\gamma^{\alpha \beta}_{[mnrpq]} D_\alpha A_\beta = 0\,,
}
which contains the linearized Dirac and Yang-Mills equations upon expanding 
in terms of $\theta$. 

Along the 
same lines, one can study the closed string or massive vertex operators and one finds the 
complete correct spectrum of the open or closed superstring. Other interesting cases one 
might study are the superstring in lower dimensions, or a finitely reducible gauge theory. 
 
The notion that one must restrict the space of the vertex operators is not new 
by itself: in the spinning (RNS) string, one should restrict the commuting susy ghosts 
to non-negative mode numbers 
\cite{big}, and also in the bosonic string one has the condition 
that vertex operators are annihilated by $b_0$ (where $b_0$ belongs to $b_{zz}$). 
We have shown \GrassiSP~that the concept of grading is nothing else that 
the ``pure ghost number'' of homological perturbation theory 
\cite{homo}. So there is, after all, a deeper geometrical meaning to the ideas we have developed. 

\section{Grading, reducibility, homological perturbation theory and BRST nilpotency}

In the previous section we have introduced a new quantum number for fields, the grading, 
and a new definition of physical states which required that vertex operators have non-negative 
grading. The results (the correct physical  spectrum) justify to some extent this notion of a grading. 
We now present a new understanding: 
{\it the grading number is the pure ghost number (resolution degree) of homological 
perturbation theory}. 

According to homological perturbation theory (HPT) 
\homo, once one has an initial BRST-like symmetry $s_0$ which 
is nilpotent modulo constraints $G_a$ and gauge transformations (with 
possibly field dependent parameter $\e^a$ where the gauge transformations 
are due to OPE's of the fields with the constraints), we may introduce new 
fields ${\cal P}_a$ and a new 
nilpotent operator\foot{This operator is known in the literature as the Koszul-Tate 
resolutor.} $\d_{-1}$ such that 
\ahin{resA}{
\delta_{-1}  {\cal P}_a = - G_a\,, ~~~~~~~ \delta_{-1} ({\rm other~fields}) = 0\,.
}
The new fields ${\cal P}_a$ carry a new quantum number usually called 
antifield number and the operator $\d_{-1}$ lowers this number. The solutions of $\delta_{-1} X =0$, 
but $X \neq \delta_{-1} Y$ are called homology instead cohomology classes because of this lowering. 
Next one relaxes the constraints $G_a= 0$ and if there is nontrivial homology, 
one introduces a new ghost which removes this spurious homology. 
In our case $G_a = \l \g^m \l$, and there is a new homology, namely $\xi^m (\g_m \l)_\a$. 
Indeed $\delta_{-1} (\xi^m (\g_m \l)_\a) = 0$, but $\xi^m (\g_m \l)_\a \neq \delta_{-1} X$ because only 
$\xi^m$ transforms under $\delta_{-1}$. Thus we add a ghost $\chi_\a$ and define 
$\delta_{-1} \chi_\a = \xi^m (\g_m \l)_\a$. Then $\xi^m (\g_m \l)_\a$ becomes trivial 
homology. Now we repeat the argument. There is again a new homological class; 
it is given by $\xi^m \xi^n + \l \g^{mn} \chi$. Indeed, 
$\delta_{-1} (\xi^m \xi^n + \l \g^{mn} \chi) = \l \g^{[m} \l \xi^{n]} + 
\l \g^{mn} ( \xi_p \g^p \l)$ and $\l \g^{mn} \g^p \l = \delta^{[n}_p \l \g^{m]} \l$. 
Again, HPT would instruct us to introduce new ghosts $B^{mn}$ and define 
$\delta_{-1} B^{mn} = \xi^m \xi^n + \l \g^{mn} \chi$. This is the conventional path. 
However, we followed another path. Namely, we introduced an antighost $b$ with 
antifield number $-3$ and this removes the extra homology class 
$ \xi^m \xi^n + \l \g^{mn} \chi$ because it is now equal to 
$\delta_{-1}b \Big( \xi^m \xi^n + \l \g^{mn} \chi  \Big)$ if at the same time we add a term 
$\oint c_z$ to $\delta_{-1}$. At this point we have a nilpotent $\delta_{-1}$ 
without any non-trivial homology classes. 

Note that we only introduced $b$ at the level 
of $\xi^m \xi^n$. We could have introduced $b$ one step earlier, namely when we removed 
$\xi^m (\g_m\l)_\a$; in that case we would not have needed a ghost $\chi_\a$ (which 
is however useful for the central charge) and still we would have obtained a nilpotent $\delta_{-1}$ 
without any non-trivial homology. However, we could not have introduced $b$ at the very beginning 
when we had the non-trivial homology  $\l \g^m \l$, because we would have gotten a trivial spectrum of the BRST 
charge.  

It has been proven that one can always add further operators $s_1, s_2, \dots$ to 
$\d_{-1} + s_0$ such that $s \equiv \d_{-1} + s_0 + s_1 + s_2 + \dots $ is nilpotent. 
The form of $s_n$ with $n > 1$ follows from the requirement that $s^2 =0$. 
The $s_n$'s have definite antifield number equal to $n$. In addition, 
one can define a further quantum number by a linear combination of the antifield 
number and the ghost number; it turns out that the {\it pure ghost number} $n_{pg}$, 
defined as the sum of the ghost number $n_g$ plus the antifield number $n_{af}$, coincides 
with our grading number. 

In the superstring case, $s_0$ should be identified with Berkovits' BRST-like symmetry in (\ref{BRSTA})
which acts on the fields $\Phi = ( \t^\a, x^m, d_\a, \l^\a, w_\a)$, where $w_\a$ is the conjugate 
momentum of $\l^\a$. This $s_0$ should be nilpotent 
up to the pure spinor constraint $\l \g^m \l =0$ and up to the gauge 
transformations $\Delta_{\e} \Phi (w) = \oint dz (\e_m \l \g^m \l) (z) \Phi(w)$ (the bracket 
$[\Phi, G_a]$ is in our case written in terms of the operator product). Indeed, 
\ahin{SSb}{
Q^2_B x^m = -{1\over 2} \l \g^m \l \,, ~~~~ 
Q^2_B \t^\a = 0\,, ~~~~ 
Q^2_B d_\a = -{1\over 2} \p (\l \g^m \l) \,,} ~~~ 
$$ 
Q^2_B \l^\a = 0\,, ~~~~
Q^2_B w_\a = \Pi_m (\g^m \l)_\a  = \Delta_{\Pi} w_\a\,.
$$
The field ${\cal P}_a$  corresponds in our case to $\xi_m$ and 
$\d_{-1} {\cal P}_a = - G_a$ corresponds to $\{Q_0, \xi^m\} = -{1\over 2} \l \g^m \l$  
where $Q_0 = \oint dz j^{B,(0)}_z(z)$ is the grading zero part of the 
BRST charge. Further, we identify $Q_2 \equiv s_1$, 
$Q_3 =s_2$ and $Q_4=s_3$ where $Q_2,Q_3$ and $Q_4$. 

The fields $\Phi$ have by definition vanishing antifield number. Hence 
$n_{pg}(\l) = n_g(\l) + n_{af}(\l) = 1 + 0 = 1$ which agrees (up to a factor 2) with our 
grading. Similarly also for all other fields $n_{pg}$  is equal to twice our grading. 
Hence, our notion of grading is closely related to the notion of antifield number in homological 
perturbation theory. 

There is however, a difference between our approach and standard homological 
perturbation theory. In the latter case one has by definition only fields with 
positive antifield number (contributing to $s_n$ with $n\geq 0$), but in our case we have antighosts 
in the theory, and if the ghosts have positive antifield number, it is reasonable to assign 
the opposite (negative) antifield number to the antighosts (in this way the action is neutral). 
One must introduce a floor from 
which to work upwards, in a similar way as Dirac introduced the concept of 
a sea to excluded unbounded negative energy. We have constructed such a floor by hand, by requiring 
that the vertex operators have a lower bound on their grading; from 
the previous correspondence it even follows that this lower bound is zero. 

We end with some comments on the previous discussion. In any application of HPT one can distinguish 
the following aspects
\begin{enumerate}
\item
the constraints one starts with may be reducible of irreducible. As constraints, following 
Berkovits, we choose $\l \g^m \l =0$ because we decompose 
$ (\l^\a d_\a)(z) (\l^b d_\b)(w) \sim ( \l \g^m \l \Pi_m )/(z-w)$ into constraints 
$\l \g^m \l$ and generators $\Pi_m$. Our set of constraints is reducible because there exists 
(field dependent and in general composite) parameters $c_m$ such that $c^m \l g_m \l =0$, 
namely $c_m = (\g_m \l)$. 
\item
One either works at the classical level or at the quantum level. We have been working at the quantum 
level. 
\item
The algebra of first class constraints may contain only first order poles, or also 
second order poles in $z-w$. We did encounter second order poles, but note that 
they were not due to double contractions, but rather to derivatives of first order poles. 
\item
We deviated from the conventional HPT by introducing the antighost $b$. 
It may be that our pair $b, c_z$ has some relation to Jacobians which arise in the 
path integral treatment of WZWN models. 
\end{enumerate}

Before concluding this section, we would like to mention the work 
by Aisaka and Kazama on an extension of the pure spinor formalism \cite{extpure}. 
They factorized the pure spinor constraints into a reducible and irreducible 
part preserving the subgroup $U(5)$ of the Lorentz group 
and, following the HPT,  they are able to derive a new BRST charge 
which is nilpotent without any constraints on the ghosts. In addition, the set of new ghost fields forms 
a system with vanishing conformal charge. It would be very interesting  to compare their formalism with our results. 
A more detailed discussion of the relation between Berkovits' formalism, our 
formalism, Aisaka and Kazama's formalism, HPT and equivariant cohomology is 
in preparation \cite{new}.

\paragraph{Acknowledgments.} 
At the July 2002 string workshop in Amsterdam, P. Townsend suggested to apply our 
ideas to a simpler model, and \cite{GrassiSR} contains the result. E. Verlinde suggested 
not to short-circuit our derivation of the BRST charge by introducing the ghost pair $(c_z,b)$ 
by hand, but to go on applying our method. This indeed works and the result will be published elsewhere. 
This work was done in part at the Ecole Normale Superieure at Paris 
whose support we gratefully acknowledge. In addition, we were partly 
funded by NSF Grant PHY-0098527.

\appendix 
\section{Majorana and Weyl spinors in $D=(9,1)$.} 
 
In $D=(9,1)$ dimensions, we 
use ten real $D=(9,1)$ Dirac-matrices $\Gamma^m = \{ I \otimes (i \tau_2), 
\sigma^\m \otimes \tau_1, \chi \otimes \tau_1\}$ 
where $m=0,\dots,9$ and $\m = 1,\dots,8$. The $\sigma^\m$ are eight real symmetric $16 \times 16$ 
off-diagonal Dirac matrices for $D= (8,0)$, while $\chi$ is the real 
$16 \times 16$ diagonal chirality matrix in $D=8$\foot{The 8 real $16 \times 16$  matrices 
of $D=(8,0)$ can be obtained from a set of 7 purely imaginary $8 \times 8$ 
matrices $\l^i$ for $D=(7,0)$ as follows $\sigma^\m = \{ \l^i \otimes \s_2, I_{8\times8} \otimes \s_1\}$. The 
seven $8 \times 8$ matrices $\l^i$ themselves can be obtained from the 
representation $\g^k = \sigma^k \otimes \tau^2, \g^4 = 1\otimes \tau^1$, and 
$\g^5 = 1\otimes \tau^3$ for, $D=(3,1)$ with real symmetric matrices $\gamma^2,\gamma^4,\gamma^5$ and 
imaginary antisymmetric $\gamma^1,\gamma^3$ as follows 
$$ 
\l^i = \{ \g^2 \otimes \s_2, \g^4\otimes \s_2, \g^5 \otimes \s_2, 
\g^1 \otimes 1, \g^3 \otimes 1, i \, \g^2 \g^4 \g^5 \otimes \s_1, i \g^2 \g^4 \g^5 \otimes \s_3\}. 
$$ 
}. 
So $\chi=\s_1, \dots, \s_8$, $\chi^T = \chi$ and $\chi^2 = 1$. 
The chirality matrix in $D=(9,1)$ is then $I \otimes \tau_3$ and the $D=(9,1)$ 
charge conjugation matrix $C$, satisfying 
$C \, \Gamma^m = - \Gamma^{m, T} C$, is given by $C = \Gamma^0$. If one uses spinors 
$\Psi^T = ( \l_L, \zeta_R)$ with spinor indices $\l_L^\a$ and $\zeta_{R,\dot{\b}}$, 
the index structure of the Dirac matrices, the charge conjugation matrix $C$, and the chirality matrix 
$\G_\# \equiv \G^0 \G^1 \dots \G^9 = I_{16\times 16} \otimes \tau_3$ 
is as follows 
\begin{equation}\nonumber 
\Gamma^m = 
\begin{pmatrix} 
 0 & (\sigma^m)^{\a \dot{\b}} \cr 
 (\tilde\sigma^m)_{\dot{\b} \g} & 0 
\end{pmatrix} 
 \,, ~~~ 
C = 
\begin{pmatrix} 
0  & c_{\a}^{~\dot{\b}} \cr 
  c^{\dot{\b}}_{~ \g} & 0 
\end{pmatrix} 
  \,, 
\end{equation} 

\begin{equation}
\G_\# = 
\begin{pmatrix} 
 I_{16\times16} &  0 \cr 
  0 &  -I_{16\times16} 
\end{pmatrix} 
  \,, 
~ 
\Gamma^9 = 
\begin{pmatrix} 
0  & 
\begin{pmatrix} I_{\small 8 \times 8} & 0 \\
0 & -I_{\small 8 \times 8}
\end{pmatrix} \cr 
\begin{pmatrix} I_{\small 8 \times 8} & 0 \\
0 & -I_{\small 8 \times 8}
\end{pmatrix}
& 0 
\end{pmatrix} 
  \,, 
\end{equation} 
 where $\sigma^m = \{I, \sigma^\mu, \chi \}$ and $\tilde\sigma^m = \{-I, \sigma^\mu, \chi \}$. 
The matrices $c_{\a}^{~\dot{\b}}$ and $ c^{\dot{\b}}_{~ \g} $ are 
 numerically equal to $I_{16\times 16}$ and 
 $-I_{16\times 16}$, respectively. 
Thus the $\l^\a$ are chiral and the 
$\zeta_{\dot\b}$ are antichiral. This explains the spinorial index structure of the $\G^m$. 
 
In applications we need the matrices $C\G^m$ (for example in \ref{pure}). Direct 
matrix multiplication shows that $C\G^m$ is given by 
\begin{equation}\label{cacB} 
C \Gamma^m = 
\begin{pmatrix} 
(\tilde\sigma^m)_{\a {\b}} & 0\\ 
 0 & - (\sigma^m)^{\dot{\b} \dot\a} 
\end{pmatrix} 
 \equiv 
\begin{pmatrix} 
 \g^m_{\a {\b}} & 0\\ 
 0  & (\g^m)^{{\b} \a} 
\end{pmatrix} 
 \,,\end{equation} 
using 
\begin{equation}\label{cacB1} 
\Gamma^{m,T} = 
\begin{pmatrix} 
0 & (\tilde\sigma^{m,T})_{\a {\dot\b}} \\ 
  (\sigma^{m,T})^{\dot{\a} \b} & 0 
\end{pmatrix} 
 \,.
\end{equation}
We only use the real $16\times 16$ symmetric matrices $\g^m_{\a\b} = \tilde\s^m_{\a\b}$ 
and $\g^{\dot\a\dot\b}_m = - \s^{m\dot\a\dot\b}$ in the text, and we omit the dots for 
reasons we now explain. 
 
The Lorentz generators are given by 
\begin{equation} \label{cacC} 
L^{mn} = {\frac 1 2 } ( \G^m \G^n - \G^n \G^m) = 
\begin{pmatrix} 
{\frac 1 2 } \s^{m, \a\dot\b} \tilde\s^n_{\dot\b\g} 
- {m \leftrightarrow n} 
& 0 
\\ 
0 & {\frac 1 2 } \tilde\s^{m}_{\dot\a \b} \s^{n,\b\dot\g} - 
{m \leftrightarrow n} 
\end{pmatrix} 
\end{equation} 
Hence the chiral spinors $\l^\a$ and the antichiral $\zeta_{\dot\b}$ form separate representation 
for $SO(9,1)$. These representations are inequivalent because $\s^m$ and $\tilde\s^m$ are equal 
except for $m=0$ where $\s^0=I$ but $\tilde\s^0 =-I$, and there is no matrix $S$ satisfying 
$S \s^\mu = - \s^\m S$ and $S \chi = - \chi S$. (From $S \s^\mu = - \s^\m S$ it follows 
that $S \chi = + \chi S$). We denote these real inequivalent representation by $\underline{16}$ 
and $\underline{16}'$, respectively. 
 
In $D=(9,1)$ dimensions one cannot raise or lower spinor indices with the charge 
conjugation matrix, because $C$ is off-diagonal. In $D=(3,1)$, on the other hand, $C$ 
is diagonal and is given by 
$C = 
\begin{pmatrix} 
\e_{\a {\b}} & 0\cr 
 0 & \e^{\dot{\a} \dot\b} 
\end{pmatrix} 
 $, and therefore one can raise and lower the indices 
with the charge conjugation matrices $\e^{\a\b}, \e_{\a\b}$ and $\e^{\dot\a \dot\b}, 
\e_{\dot\a\dot\b}$. For that 
reason one has in  $D=(3,1)$ two independent representations: 
 $\l^\a \sim \l_\a$ and $\chi_{\dot\b}  \sim \chi^{\dot\b}$.

In $D=(9,1)$ dimensions, one can also define spinors $\k_\a$ and $\eta^{\dot \a}$ which 
transform under Lorentz transformations such that $\kappa_\a \lambda^\a$ and 
$\eta^{\dot\alpha} \chi_{\dot\alpha}$ are invariant. If we denote the generators of $\l^\a$ by 
$(\g^{kl}, \g^k)$ with $k,l=1,\dots,8$,  those for $\chi_{\dot \a}$ are given by 
$(-\g^{kl,T}, -\g^{k,T} )$. Of course these matrices form also a representation 
of the Lorentz group, but they are not inequivalent representations. It is easy to check that in the representation 
given above, the Lorentz generators for the spinors $\l^\a,\chi_{\dot\a}, \kappa_\a$, and $\eta^{\dot\alpha}$ 
are given, respectively, by 
\begin{equation} \label{cacD} 
(\g^{kl}, \g^k), ~~(\g^{kl}, -\g^k), ~~(\g^{kl}, -\g^k), ~~ (\g^{kl}, \g^k)\,.
 \end{equation} 
Thus in $D=(9,1)$ dimensions $\kappa_\alpha$ transforms like $\chi_{\dot \a}$, 
and $\eta^{\dot \a}$ like $\l^\a$. Hence, one may omit the dots without causing confusion, but it matters whether one 
has upper or lower indices. For $D=(3,1)$ dimensions one has just the opposite situation: the representation 
to which $\l^\a$ and $\kappa_\a$ belong is inequivalent to the representation to which $\chi_{\dot \a}$ and 
$\eta^{\dot \a}$ belong. 

We conclude that chiral spinors are given by $\l^\a$, antichiral spinors by $\chi_\a$ and in the text we use 
the twenty real symmetric $16\times 16$ matrices $\g^m_{\a\b}$ and $\g^{m, \a\b}$ (omitting again the 
dots in the latter). The matrices $\g_m$ satisfy 
$\g^m_{\a\b} \g^{n\,\b\g}+ 
\g^n_{\a\b} \g^{m\,\b\g}=2\eta^{mn}\d_\a^\g$ and 
$\g_{_{m\,(\a\b}} \g^{_{m}}_{_{\g)\d}}=0$. The latter relation makes Fierz rearrangements very easy. 
The usual Fierz rearrangement for 3 chiral spinors becomes then simply 
the statement that $\g^{_m}_{_{\a\b}} \g_{_{m \g\d}}$ vanishes when totally symmetrized in the indices $\a,\b$ and $\g$. In particular, 
\ahin{ccccc}{
(\lambda \gamma^m \lambda) \gamma_m\lambda =0\,.}


\section{The WZNW term} 
 
We follow \cite{Mez}. 
The WZNW term ${\cal L}_{WZ}$ is proportional to $\e^{\mu\nu}$ (with 
$\mu,\nu =0,1$) hence ${\cal L}_{WZ} d^2x$ can be written as a 2-form 
\ahin{WWA}{ 
\omega_2 \equiv {\cal L}_{WZ} d^2x\,. 
} 
Since $\omega_2$ is susy invariant up to a total derivative, 
we have 
\ahin{WWB}{ 
\d_\e \omega_2 = d X\,. 
} 
 
Define now a 3-form $\omega_3$ as follows: $\omega_3 = d \omega_2$. Then clearly, 
\ahin{WWC}{ 
\d_\e \omega_3 = 0\,, ~~~~~~~ d \omega_3 = 0\,. 
} 
 
{}From $\d_\e \omega_3=0$ it is natural to try to construct 
$\omega_3$ from the susy-invariant 1-forms $\Pi^m = d x^m - i \sum_j \t^j \g^m d \t_j$ and $d \t^i$. 
Lorentz invariance then yields only one possibility 
\ahin{WWD}{ 
\omega_3 = a_{ij} \Pi^m d \t^i \g^m d \t^j\,. 
} 
where $a_{ij}$ is a real symmetric $N \times N$ matrix. We diagonalize $a_{ij}$ by a real orthogonal 
transformation (which leaves $\Pi^m$, and thus ${\cal L}_1$ in (\ref{susyC}) invariant). Then 
$d \omega_3 = i \left(\sum_i d \t^i \g^m d \t^i \right) \left( \sum_k a_k d \t^k \g^m d \t^k  \right)$. 
In $d \omega_3$ the direct terms cancel due to the 
standard identity $\g^m d\t^1 (d\t^1 \g_m d\t^1) =0$, while the cross-terms cancel only if 
$N=2$ and if the diagonal matrix $a_{ij}$ has entries $(+1,-1)$. Hence 
\ahin{WWD_}{ 
\omega_3 =  i \Pi^m \left( d \t^1 \g_m d \t^1 - d\t^2 \g^m d\t^2    \right)\,. 
} 
Using that $\omega_3 = d \omega_2$, 
we find the WZNW term up to an overall constant 
\ahin{WWE}{ 
{\cal L}_{WZ} =  {\frac 1 \pi} \e^{\m\n} 
\left[ - i 
\p_\m x^m ( \t^1 \g_m \p_\n \t^1 - \t^2 \g_m \p_\n \t^2 ) + \t^1 \g_m \p_\m \t^1  \t^2 \g^m \p_\n \t^2 
\right]\,. 
} 
Indeed, with $\e^{\mu\nu} d^2x = dx^\mu dx^\nu$ one gets
\begin{eqnarray}\label{WWF} 
d ({\cal L}_{WZ} d^2x) & \sim &- i d x^m 
\left( d \t^1 \g_m d \t^1 - d\t^2 \g^m d\t^2 \right) 
\nonumber 
\\ 
& + & 
\left(\t^1 \g_m d \t^1\, d\t^2 \g_m d\t^2 - d \t^1 \g_m d \t^1\, \t^2 \g^m d\t^2 
\right) 
\end{eqnarray} 
which is equal to 
\ahin{WWG}{ 
\omega_3 = - i \left(d x^m - \t^1\g_m d \t^1 - \t^2\g^m d\t^2 \right) \left( d\t^1\g_m d\t^1 - d\t^2\g_m d\t^2 \right)\,. 
} 
 
Note that the WZNW term is antisymmetric in $\t^1$ and $\t^2$ while ${\cal L}_1$ 
is symmetric. Only the sum of ${\cal L}_1$ and ${\cal L}_{WZ}$ is $\k$-invariant, up to a total 
derivative. The $\kappa$-transformation rule for $x^m$ is $\d_\k x^m = - \sum_j \d_\k \t^j 
\g^m \d_\k \t^j$ with the opposite 
sign to the susy rule. The expressions 
for $\d_\k \t^\a$ and $\d_\k \sqrt{-h} h^{\m\nu}$ are complicated, involving self-dual 
and antiselfdual anticommuting gauge parameters with 3 indices, but we do not need them. We begin 
with the BRST law $s\, \t^\a = i \l^\a$ where $\l^\a$ is an unconstrained ghost field, but the precise 
classical action to which this corresponds is not know at the present. That does not matter as long as 
we can construct the complete quantum theory, although knowledge of the classical action might clarify the 
results obtained at the quantum level. 
 
For the open string one has the following boundary conditions at $\s=0,\pi$ \GrassiUG 
\ahin{WWH}{ 
\t^{1 i} = \t^{2 i}\,, ~~~~ 
\e^{1 i} = \e^{2 i}\,, ~~~~ 
h^{\s \b} \p_\b x^m =0\,, ~~~~ 
\k^{1}_t = \k^{2}_t = \sqrt{-h} \k^{1\sigma} = \sqrt{-h} \kappa^{2\sigma}\,. 
} 
 
 
\section{A useful identity for the superstring} 
 
The superstring action is given by 
\ahin{GS}{ 
{\cal L} = - \frac{ 1}{ 2  \pi}  \, \eta_{mn} \, \Pi^m_\mu \Pi^{n\m}  + {\cal L}_{WZ} \,, 
} 
$$ 
{\cal L}_{WZ} = {\frac 1 \pi} \e^{\m\n} 
\left[- i 
\p_\m x^m ( \t^1 \g_m \p_\n \t^1 - \t^2 \g_m \p_\n \t^2 ) + \t^1 \g_m \p_\m \t^1  \t^2 \g^m \p_\n \t^2 
\right] 
$$ 
where $\Pi^m_\mu$ is given in \susyB. 
For definiteness we choose $\e^{01} = 1$ and $\eta^{\m\n}$ as well as 
$\eta^{mn}$ have $\eta^{00} = -1$. This action is real. 
 
By just writing out all the terms, the action can be re-written with chiral 
derivatives 
\ahin{GSA}{ 
- \pi {\cal L} = \eta_{mn} \, \p x^m \bar\p x^n 
- i \p x^m  \t^1 \g_m \bar\p\t^1 
- i \bar\p x^m  \t^2 \g_m \p\t^2 
} 
$$ 
- {\frac 1 2 } (\t^1 \g^m \bar\p \t^1)( \t^1 \g_m \p \t^1 + \t^2 \g_m \p \t^2) 
- {\frac 1 2 } (\t^2 \g^m \p \t^2)( \t^1 \g_m \bar\p \t^1 + \t^2 \g_m \bar\p \t^2) 
$$ 
with $\p = \p_\s - \p_t$ and $\bar\p = \p_\s + \p_t$. 
 
Except for the purely bosonic terms, all terms involve either $\bar\p\t^1$ or $\p\t^2$. Hence 
we can write the action as 
\ahin{GSB}{ 
- \pi {\cal L} = \frac{1}{2} \eta_{mn} \, \p x^m \bar\p x^n + (p_{1 \a})_{\rm Sol}  \bar \p \,\t^{1 \a} + 
(p_{2 \a})_{\rm Sol} \p \, \t^{2\a} 
} 
where $(p_{i \a})_{\rm Sol}$ are complicated composite expressions. 
 
We can then 
also write the action with independent $p_{i \a}$ if we impose the constraint that 
$d_{i \a}  \equiv p_{i \a} - (p_{i \a})_{\rm Sol}$ vanishes. 
The complete expressions for $d_{j\a}$ are given by 
\ahin{GSC}{ 
d_{1 \a} = p_{1 \a} +  i \p x^m \g_m \t^1 + 
{\frac 1 2 } (\g^m \t^1)( \t^1 \g_m \p \t^1 + \t^2 \g_m \p \t^2) \,, 
} 
$$ 
d_{2 \a} = p_{2 \a} +  i \bar\p x^m  \g_m \t^2 + 
{\frac 1 2 } (\g^m \t^2)( \t^1 \g_m \bar\p \t^1 + \t^2 \g_m \bar\p \t^2) \,. 
$$ 
In the text we work with the free action with independent fields $p_{i \a}$. The 
$d_{i \a}$ are transferred to the BRST charge where they are multiplied by the independent 
unconstrained real chiral commuting spinors $\l^\a$.  To make use of the calculation technique 
of conformal field theory, we make a Wick rotation $t \rightarrow - i \tau$, $\p_t \rightarrow 
+ i \p_\tau$ and $\p = \p_\s - \p_\tau \rightarrow \p = \p_\s - i \p_\tau$ and analogously for $\bar\p$. 
We also restrict ourselves to only one sector with $\t = \t^1$ and $d_{\a} = d_{1\a}$, by setting $\t^2 = 0$. 
For a treatment which describes both sectors, we refer to \GrassiXF. 
 
 
\section{Solution of the pure spinor constraints.} 
 
In this appendix we discuss a solution of the constraint that the 
chiral spinors $\l$ are pure spinors. The equation to be solved 
reads 
\ahin{puu}{ 
\l^\a \g^m_{\a\b} \l^\b =0\,, 
} 
where $\l^\a$ are complex chiral (16-component) spinors. We shall decompose $\l$ w.r.t. a non-compact 
version of a $SU(5)$ subgroup of $SO(9,1)$ as 
$|\l\rangle = \l_+ |0\rangle + {\frac {1}{ 2 !}} \l_{ij} a^i a^j |0\rangle +  {\frac {1}{ 4 !}} \l_{jklm} a^j a^k  a^l a^m |0\rangle$. 
This decomposition corresponds to $\underline{16} = \underline{1} + \underline{10} + \underline{5}^*$. Then 
we shall show that the constraints express the $\underline{5}^*$ in terms of the $\underline{1}$ and $\underline{10}$. 
Hence there are 11 independent complex components in $\l$. We shall prove that $\l$ is complex and not a Majorana 
spinor, so $\bar\l_D \equiv \l^\dagger i \g^0$ differs from $\bar\l_M = \l^T C$. 
(Recall that a Majorana spinor is defined by the 
condition $\bar\l_D = \bar\l_M$). 
 
The Dirac matrices in $D=(9,1)$ dimensions satisfy $\{\G^m, \G^n\} = 2 \eta^{mn}$, where $\eta^{mn}$ is 
diagonal with entries $(-1,+1, \dots, 1)$ for $m,n=0,\dots,9$. We combine them into 5 annihilation operators $a_j$ 
and 5 creation operators $a^j = a^\dagger_j$ as follows 
\ahin{puuA}{ 
a_1 = \half(\G^1 + i \G^2)\,, ~~~~~ 
a_2 = \half(\G^3 + i \G^4)\,, ~~~\dots~~~~ 
a_5 = \half(\G^9 -  \G^0)\,. 
} 
$$ 
a^1 = \half(\G^1 - i \G^2)\,, ~~~~~ 
a^2 = \half(\G^3 - i \G^4)\,, ~~~\dots~~~~ 
a^5 = \half(\G^9  +  \G^0)\,. 
$$ 
Clearly $\{ a_i, a^j \} = \d_i^j$ for $i,j= 1,\dots,5$. We introduce a vacuum $|0 \rangle$ 
with $a_i |0 \rangle = 0$. By acting with one or more $a^j$ on  $|0 \rangle$, we obtain 32 states $| A \rangle$ 
with $A =1, \dots, 32$. . 
Similarly, we introduce a state $\langle 0|$ which satisfies  $\langle 0|a^j = 0 $ and we create 32 states $\langle B|$ by acting 
with one or more $a_i$ on $\langle 0|$. We choose the states $\langle B|$ as $| A \rangle^\dagger$. For example, 
if $| A \rangle = a^{i_1} \dots a^{i_k} |0 \rangle $ then $\langle A| = \langle 0| a_{i_k} \dots a_{i_1}$. Then 
$\langle A | B \rangle = \delta^A_{~~B}$.

{\it Lemma 1:} The matrix elements $\langle B| a^j |A\rangle \equiv (\G^j)^B_{~~A}$ and 
$\langle B| a_j |C\rangle \equiv (\G_j)^B_{~~C}$ form a representation 
of the Clifford algebra. 
 
{\it Proof:} This follows from $\sum |C\rangle  \langle C| = I$. Namely, 
$\sum |C\rangle  \langle C|  =  |0 \rangle  \langle 0| + \sum_i a^i  |0 \rangle  \langle 0| a_i + \dots + 
 a^1\dots a^5  |0 \rangle  \langle 0| a_5 \dots a_1$, where the sum over $C$ runs over the 32 states shown. 
For any state $| A \rangle$ one has $| A \rangle = \sum |C\rangle  \langle C| A \rangle$, because 
$  \langle C| A \rangle = \d^C_{~~A}$ by construction. 
 
{\it Lemma 2:} The chirality matrix $\G_\# = \G^1\G^2\dots\G^9\G^0$ satisfies $\G_\#^2 = 1$, and 
$\G_\#^\dagger = \G_\#$. It is given by 
\ahin{puuB}{ 
\G_\# = (2 a_1 a^1 -1) \dots ( 2 a_5 a^5 -1) \,. 
} 
 
{\it Proof:} $\G_\# = (a^1 + a_1)(a^1 - a_1) \dots (a^5 + a_5) (a^5 - a_5)$ and 
$ (a^1 + a_1)(a^1 - a_1) = (2 a_1 a^1 -1)$. As a check note that $(2 a_1 a^1 -1)^2=1$, and that 
$\{\G_\#, a^1 \} =0$ because $\{  (2 a_1 a^1 -1) , a^1 \} =0$. Similarly $\{\G_\#, a_1\} =0$. Further, 
$\G_\# |0 \rangle =  |0 \rangle$. 
 
{\it Lemma 3:} $\langle B| a^j |C\rangle = \langle C| a_j |B\rangle =$ real. 
 
{\it Proof:} This follows from the fact that one obtains the second matrix element 
from the first by left-right reflection, and from the fact that the anticommutation 
relations have the same symmetry and are real: $\{a^k, a_l\} =\{a_l, a^k\} = \d^k_{~l}$. 
 
{\it Lemma 4:} The matrix representation of $\G^1,\G^3,\G^5,\G^7, \G^9$ is real 
and symmetric while that of $\G^2,\G^4,\G^6, \G^8$ is purely imaginary and antisymmetric, 
and that of $\G^0$ is real and antisymmetric. 
 
{\it Proof:} $\langle A| a^j \pm a_j |B \rangle = \langle B| \pm a^j + a_j |A\rangle$. 
 
{\it Lemma 5:} The charge conjugation matrix $C$, defined by $C \G^m = - \G^{m,T} C$ is given by 
$C = - \G^2 \G^4 \G^6 \G^8 \G^0 = (a_1 - a^1) (a_2 -a^2) \dots (a_5 -a^5)$. The minus sign is added for later 
convenience. 
 
{\it Proof:} $\G^1,\G^3,\G^5,\G^7, \G^9$ anticommute with $C$, while 
$\G^2,\G^4,\G^6, \G^8,\G^0$ commute with $C$, the former are symmetric while the latter are antisymmetric. 
 
{\it Theorem I:} A chiral spinor $\l$ can be expanded as follows 
\ahin{puuD}{ 
|\l \rangle = \l_+  |0\rangle + 
{\frac 1 {2 !}} \l_{ij} a^j a^i |0\rangle +  {\frac 1 {4 !}} \l^i \e_{ijklm} a^j a^k  a^l a^m |0\rangle\,. 
} 
 
{\it Proof:} $\G_\# |0\rangle = |0\rangle$; hence $\G_\# |\l \rangle = |\l \rangle$. The 16 non-vanishing 
components of $|\l \rangle$ are the projections of the ket $|\l \rangle$ onto the corresponding 16 bras: 
in particular 
\ahin{puuE}{ 
\l_+ = \langle 0| \l \rangle = \langle \l|0 \rangle \,, 
~~~~~ 
\l_{ij} = {\frac 1 {2 !}} \langle 0|a_i a_j| \l \rangle = {\frac 1 {2 !}} 
\langle \l| a^j a^i|0 \rangle \,, 
} 
$$ 
 \l^i =  {\frac 1 {4 !}} \e^{ijklm} \langle 0| a_j a_k  a_l a_m | \l \rangle = 
{\frac 1 {4 !}} \e^{ijklm} \langle \l| a^j a^k  a^l a^m |  0 \rangle\,. 
$$ 
 
We are now ready to solve the ten constraints $\bar\l^\a \G^m_{\a\b} \l^b =0$. These 
relations are equivalent to the five constraints $\l^T C a^j \l = 0$ and the five other 
constraints $\l^T C a_j \l = 0$. They can be rewritten as follows 
\ahin{puuF}{ 
\langle \l| C a^j | \l \rangle = 0\,, ~~~~~ 
\langle \l| C a_j | \l \rangle \,. 
} 
 
{\it Theorem II:} $\langle A| C |B\rangle \neq 0$ iff $A^\dagger B$ is 
proportional to precisely $a^1a^2 a^3 a^4 a^5$. 
 
{\it Proof:} $a_j C = - C a^j$ and $a^j C = - C a_j $, Further $C |0\rangle = - a^1a^2 a^3 a^4 a^5 |0\rangle$ 
and $\langle 0| C =  \langle 0|  a^5a^4 a^3 a^2 a^1$. Pulling all $a_j$ in $\langle A|$ to the right of 
$C$, we obtain, up to an overall sign, $\langle 0| C A^\dagger B |0 \rangle$ and this is only non-vanishing if all $a^k$ in 
$A^\dagger B$ match the $a_k$ in $\langle 0|C$. It follows that $ \langle 0| C a^1a^2 a^3 a^4 a^5 |0\rangle = 1$. 
 
{\it First set of constraints} 
\ahin{puuG}{ 
\langle \l| C a^{i_0} | \l \rangle = \langle 0|C \left( \l_+ + {\frac 1 2 } \l_{ij} a^i a^j + 
{\frac 1 {4 !}} \l^i a^j a^k a^l a^m \e_{ijklm} \right) a^{i_0} \l \rangle 
 } 
$$ 
=2 \left( \l_+ \l^{i_0} + {\frac 1 {4 !}}  \e^{i_0 jklm} \l_{jk} \l_{lm} \right)\,. 
$$ 
 
{\it Second set of constraints} 
\ahin{puuH}{ 
\langle \l| C a_{i_0} | \l \rangle = \langle 0|C \left( \l_+ + {\frac 1 2 } \l_{ij} a^i a^j + 
{\frac 1 {4 !}} \l^i a^j a^k a^l a^m \e_{ijklm} \right) a_{i_0} \l \rangle 
 } 
$$ 
=- 2 \l_{i_0 j} \l^j \,. 
$$ 
 
{\it Main Result:} The solution of the first set of constraints 
$\l_+ \l^{i} + {\frac 1 {4 !}}  \e^{i jklm} \l_{jk} \l_{lm} = 0$ is given by 
\ahin{puuI}{ 
\l^i = - {\frac 1 {4 ! \l_+}}  \e^{i jklm} \l_{jk} \l_{lm}\,. 
} 
The solution automatically satisfies the second set of constraints because 
\ahin{puuK}{ 
\l^i \l_{in} = \e^{ijklm} \l_{jk} \l_{lm} \l_{in} = 0\,. 
} 
 
{\it Proof:} A totally antisymmetric tensor with $6$ indices in 5 dimensions vanishes. Hence 
$\l^i \l_{in}$ is equal to a sum of 5 terms, due to exchange $n$ with $j,k,l,m$ and $i$, respectively. Interchanging 
$n$ with $i$ yields minus the original tensor, but also interchanging $n$ with $j,k,l$ and $m$ yields each time minus the original 
expression. Hence the expression vanishes. 
 
{\it Comment 1:} The fact that a pure chiral spinor contains 
11 independent complex components leads to a vanishing 
central charge in Berkovits' approach with variables $x^m, \t^\a$ and the 
conjugate momentum $p_\a$, and $\l^\a$ with conjugate momentum $p_{(\l)\a}$: 
$c = + 10_{x} - 2 \times 16_{\t p} + 2 \times 11_{\l, p_{\l}} = 0$. 
In our approach we have $16$ independent real component in $\l^\a$ and 16 conjugate 
momenta $p_{(\l)\a}$ with $\a =1,\dots,16$. Also in our case $c=0$, but there are more ghosts, and 
there is nowhere a decomposition w.r.t. a subgroup of $SO(9,1)$. 
 
{\it Comment 2:} 
In the decomposition in Theorem I, one can choose all $\l$'s to be real, and $\l^i$ to be expressed in terms 
of $\l_+$ and $\l_{ij}$ as in \puuI. Then $\l$ is a real chiral spinor. However, the Dirac matrices are complex, 
so under a Lorentz transformation $\l$ becomes complex in a general Lorentz frame. 
 
 

\end{document}